# Stakeholder Criteria in Technical Debt Decision-Making: A Practitioner-Informed Taxonomy

João Pedro Bittencourt
Federal University of Bahia
Salvador, Bahia, Brazil
joaoqueiroz@ufba.br

Rita Suzana P. Maciel
Federal University of Bahia
Salvador, Bahia, Brazil
rita.suzana@ufba.br

## Abstract

Technical Debt (TD) decisions are rarely just technical. Stakeholders consider heterogeneous criteria, such as deadlines, client expectations, architectural consequences, available resources, organizational authority, and team sustainability, when deciding whether to acquire, postpone, prioritize, or repay TD. However, these criteria are often reported under different labels and at different levels of abstraction, making it difficult to compare findings across TD decision contexts. This paper proposes a practitioner-informed taxonomy represented as a conceptual model of stakeholder criteria for TD decision-making. Based on a qualitative study with 11 software practitioners in Brazil, we extracted, consolidated, and organized criteria related to TD acquisition and repayment. The resulting taxonomy comprises six families: stakeholder-facing value; delivery and resource pressure; technical integrity and systemic risk; decision basis and epistemic style; governance and legitimation; and human and team sustainability. Our findings show that acquisition and repayment share taxonomic families but differ in decision function. In acquisition, criteria often operate as permission mechanisms that make shortcuts acceptable under current constraints. In repayment, criteria operate as authorization mechanisms that justify allocating time and resources to address existing debt. The conceptual model further shows how criteria become actionable through stakeholder interpretation, translation across technical and business perspectives, and organizational legitimation. The taxonomy and model provide an empirically grounded artifact for classifying TD decision criteria, comparing decision contexts, and structuring discussions about TD acquisition and repayment.

## 1 Introduction

Technical Debt (TD) refers to short-term compromises in software development that may incur future costs [6, 13]. Although the metaphor is often associated with technical shortcuts, TD decisions are rarely made solely on technical grounds. In practice, stakeholders must weigh client expectations, deadlines, available resources, architectural consequences, organizational constraints, and the perceived legitimacy of spending time on non-feature work [2, 13, 22, 25].

This heterogeneity challenges both research and practice. Prior TD research has identified several factors influencing TD decisions, including time pressure, cost, risk, customer satisfaction, maintainability, and organizational constraints [9, 14, 22]. Yet, these factors are reported under varied labels and levels of abstraction. For example, stakeholder-related concerns may appear as client satisfaction, user value, business value, or stakeholder priority, while time may refer to deadline pressure, time-to-market, future effort, or operational slack [2, 9, 14, 22]. This fragmentation hinders the comparison of empirical results and obscures how similar criteria function across TD decision contexts [11].

Existing TD taxonomies help organize the nature of debt items, such as code, architecture, or test debt [1, 15, 22], whereas other empirical artifacts have mapped TD management practices and reasons for non-payment [9]. These contributions are essential, but they address a different problem: organizing debt by type or management moment rather than organizing the criteria stakeholders use when reasoning about TD decisions. This distinction matters because TD management depends not only on the kind of debt involved, but also on how stakeholders justify, delay, negotiate, or authorize action.

This paper addresses this gap by proposing a practitioner-informed taxonomy represented as a conceptual model of stakeholder criteria for TD decision-making. Rather than offering another classification of TD types or a predictive model, the proposed artifact organizes decision criteria into families and shows how they are mobilized, interpreted, translated, and legitimized across two contrasting TD decision moments: acquisition and repayment.

We focus on acquisition and repayment because they are related but asymmetric [4, 16]. Acquisition involves legitimizing a compromise under current constraints, whereas repayment involves justifying time and resources to address an existing compromise. The same criterion can appear in both moments but serve different functions. For instance, time pressure can justify acquiring TD when a deadline is approaching, while lack of time can postpone repayment [23]. Thus, understanding TD decisions requires examining not only which criteria are involved, but also what those criteria do in the decision process [5].

We conduct a qualitative study with 11 software practitioners in Brazil. From interview analysis, we extracted and organized stakeholder criteria related to TD acquisition and repayment [2]. We then refined the resulting families through a cross-decision stress test and a literature-informed cross-check with prior TD research. We address the following research questions:

- **RQ1.** What stakeholder criteria appear across TD decision-making contexts?
- **RQ2.** How can these criteria be organized into a practitioner-informed taxonomy represented as a conceptual model?
- **RQ3.** What structural asymmetries does the taxonomy reveal between criteria used in TD acquisition and repayment?

Our findings organize TD decision criteria into six taxonomic families that combine technical concerns with stakeholder value,

delivery and resource pressure, epistemic basis, governance, and human consequences. The conceptual model further shows that these criteria become actionable through stakeholder interpretation, translation across technical and business perspectives, and organizational legitimation.

As a highlight, while acquisition and repayment share taxonomic families, these families play different roles in the decision process. In acquisition, criteria often act as *permission mechanisms*, justifying shortcuts under current constraints. In repayment, they operate as *authorization mechanisms*, justifying the allocation of present time and resources to existing debt. This distinction helps explain why TD is often easier to acquire than to repay.

This paper outlines three main contributions. First, it provides a qualitative, empirically grounded taxonomy of stakeholder criteria for TD decisions. Second, it represents this taxonomy as a conceptual model that explains how criteria are mobilized and legitimized across acquisition and repayment contexts. Third, it identifies the functional asymmetry of these criteria, offering an analytical artifact for researchers and practitioners to structure TD discussions and compare decision contexts.

The remainder of this paper is organized as follows. Section 2 reviews background and related work. Section 3 describes the research method. Section 4 presents the findings, taxonomy, and conceptual model. Section 5 discusses the findings and implications. Section 6 examines threats to validity, and Section 7 concludes.

## 2 Technical Debt Decision-Making

The TD metaphor explains how short-term development compromises lead to future costs, much like financial debt and interest [6, 13]. Since then, TD literature has expanded to include debt types, management, prioritization, measurement, and repayment [14, 15, 22]. Much of this work focuses on technical and economic aspects, such as identifying debt, estimating its cost, prioritizing repayment, and integrating TD management into software processes [12, 19].

Secondary and tertiary studies have helped consolidate this body of knowledge by organizing TD types, management activities, tools, strategies, and practitioner-oriented information [15, 22]. These studies reduce conceptual dispersion in a field where the TD metaphor has been used to describe heterogeneous technical, process, and organizational phenomena.

However, TD decisions are not only technical or economic. Empirical studies show that TD management is shaped by communication, team coordination, organizational constraints, stakeholder expectations, deadlines, and value conflicts [2, 13, 22, 25]. Practitioners may knowingly incur TD to meet deadlines, satisfy clients, preserve momentum, or reduce coordination costs, and may postpone repayment even knowing the technical consequences. Repayment may lack visible business value, compete with feature delivery, or require organizational approval [9, 19, 24]. These findings suggest TD decision-making is a socio-technical activity involving multiple stakeholder perspectives and interpretations of value.

This socio-technical perspective matters because technical and business-oriented stakeholders often participate in the same decision process while emphasizing different concerns. Technical stakeholders may focus on maintainability, architecture, testing, or system evolution, whereas business-oriented stakeholders may prioritize delivery commitments, customer satisfaction, time-to-market, or budget constraints [2, 13, 19]. Thus, TD decisions require not only identifying technical consequences but also negotiating which consequences matter, to whom, and under what conditions [2, 22, 25].

Systematic reviews and empirical studies often list elements that influence TD decisions, such as cost, time-to-market, risk, customer satisfaction, maintainability, productivity, team motivation, and organizational culture [9, 14, 22, 26]. However, these elements are represented inconsistently: some studies treat them as decision criteria [2, 14, 21], while others classify them as causes, contextual factors, motivations, consequences, or prioritization inputs [7, 14]. This variation creates analytical fragmentation. Without a systematic organization of decision drivers, it is difficult to compare studies or determine whether different terms refer to distinct phenomena or to the same decision logic.

This creates a methodological challenge for empirical TD research. If decision drivers are reported as flat lists of factors, researchers may lose sight of the roles those factors play in decisions. For example, time may refer to deadline pressure, time-to-market, development speed, future effort, or repayment slack. Similarly, quality may refer to user-visible product quality, internal code quality, process quality, architecture, or maintainability. A useful classification must therefore distinguish both the criteria present and their decision functions.

Existing TD taxonomies have classified the nature of debt itself, including code, design, architecture, test, documentation, and requirements debt [1, 15, 22]. Other artifacts organize specific TD management concerns. For example, Freire et al. [9] mapped TD payment practices, reasons for non-payment, decision factors, and impediments. Freire et al. [10] organized TD management practices, impediments, decision factors, enabling practices, and actions into IDEA diagrams, while Mandic et al. [17] proposed prioritization schemas distinguishing technical and nontechnical perspectives. These contributions show the usefulness of organizing TD-related factors into models, diagrams, or schemas, but they do not provide a cross-decision organization of stakeholder criteria for both acquiring and repaying TD.

This distinction motivates our work: rather than proposing another taxonomy of TD types, we focus on stakeholder criteria for TD decision-making. Prior work shows that TD management decisions depend on decision-making factors, practices, techniques, and stakeholder viewpoints [8]. Building on this, our taxonomy structures stakeholder criteria to compare how similar criteria function across TD acquisition and repayment.

We use taxonomy as an empirical and conceptual artifact that supports classification, comparison, and traceability [3, 20]. It does not predict decisions or prescribe a universal process; instead, it provides stable categories for naming criteria, distinguishing related factors, and comparing their functions across decision contexts. Thus, our contribution is not a taxonomy of TD artifacts or TD types, but a practitioner-informed taxonomy represented as a conceptual model of stakeholder criteria in TD decision-making.

This need is especially relevant when contrasting TD acquisition and repayment. Both are decision moments, but they are not symmetric. Acquisition legitimizes a compromise, whereas repayment justifies allocating time and resources to address it. Although the same criterion may appear in both moments, it serves different purposes. Therefore, a taxonomy of stakeholder criteria can help capture this shared vocabulary while distinguishing how criteria function throughout the TD lifecycle.

## 3 Research Method

This section describes how we constructed the taxonomy from an interview-based qualitative study with software practitioners. We present the study design, participants, analytical artifacts, taxonomy construction process, and strategies for rigor and traceability.

### 3.1 Study Design

We conducted a qualitative study on stakeholder criteria in TD decision-making, based on semi-structured interviews with 11 software practitioners in Brazil. Participants held both technical and business-oriented roles, including project managers, requirements analysts, business analysts, developers, and technical leaders. This range of profiles enabled us to examine TD decision-making from multiple stakeholder perspectives.

The empirical material used in this paper comes from a broader interview study on how practitioners reason about Technical Debt (TD) decisions in real software development contexts. Previous work [2] analyzed part of this material with a focus on TD acquisition. In contrast, this paper reuses the interview-derived analytical artifacts to identify, consolidate, and organize stakeholder criteria across both TD acquisition and repayment contexts.

The interview protocol investigated how practitioners reason about TD acquisition and repayment. In this article, we focus specifically on how stakeholders justify, postpone, negotiate, or authorize TD-related decisions. The unit of analysis is the stakeholder criterion, defined as a named concern, consideration, or evaluative element mobilized by practitioners when reasoning about TD.

Our objective is to develop an empirically grounded taxonomy of stakeholder criteria in TD decision-making. TD acquisition and repayment are treated not as the primary objects of analysis, but as complementary decision contexts that reveal how analogous criteria may be applied or interpreted differently. This positions the study as primarily taxonomic, with a secondary comparative perspective.

### 3.2 Participants and Data Collection

We recruited participants from the Brazilian software industry using convenience and professional network sampling. All participants worked in Brazil during data collection. The sample comprised experienced professionals: seven reported over 10 years in software development, three had 5–10 years, and one had less than 2 years. Their experience spanned agile, hybrid, and plan-driven practices. Table 1 summarizes the participants' roles, experience, and main sector or context.

The interviews followed a semi-structured format. The interview schedule covered participants' roles, responsibilities, development context, familiarity with TD, concrete examples of TD, criteria used when acquiring or repaying TD, and stakeholder interaction during TD-related decisions. It also included recurring decision dimensions such as cost, time, quality, risk, benefit, and stakeholder satisfaction.

**Table 1: Participant profile.**

| Code | Role | SE experience | Context |
|---|---|---|---|
| P1 | Project Manager | 10+ years | Private sector |
| P2 | Development Analyst | 10+ years | Public sector |
| P3 | Requirements Analyst | 5–10 years | Public sector |
| P4 | Project Manager | < 2 years | Private sector |
| P5 | Senior Backend Developer | 10+ years | Private sector |
| P6 | Technical Lead | 5–10 years | Private sector |
| P7 | Business Analyst | 10+ years | Private sector |
| P8 | Requirements Analyst | 5–10 years | Public sector |
| P9 | Requirements Analyst | 10+ years | Mixed |
| P10 | Project Manager | 10+ years | Mixed |
| P11 | Developer and Technical Lead | 10+ years | Private sector |

The complete interview schedule is available in the replication package.

Participants described concrete TD decision situations involving criteria, stakeholder roles, negotiation, business pressure, technical concerns, and repayment conditions. Follow-up questions were used only to clarify or deepen participants' accounts, not to steer them toward predefined answers.

### 3.3 Analytical Artifacts

The taxonomy was constructed from four interview-derived analytical artifacts: a coding table containing code identifiers, descriptions, criteria, and traceability information; a mapping of TD acquisition criteria to patterns and themes; a mapping of TD repayment code identifiers to patterns and themes; and descriptive documents explaining the acquisition and repayment patterns.

We also used prior TD literature as a post-analysis reference for terminology refinement and conceptual positioning [10, 15, 17, 22]. This literature-informed cross-check was not used to impose categories a priori; rather, it helped examine whether the emerging taxonomy was aligned with established TD terminology and defensible in relation to prior TD research.

These artifacts preserved a chain of evidence from interview-derived codes to higher-level criteria, patterns, and themes. This traceability was essential because the taxonomy required abstraction across decision contexts while remaining anchored in the empirical material.

### 3.4 Taxonomy Process

The taxonomy construction followed a five-step process: criterion inventory, consolidation and normalization, family construction, cross-decision stress test, and literature-informed cross-check. Figure 1 summarizes these steps.

**Step 1: Criterion inventory.** We extracted named criteria from the coding table and acquisition/repayment mappings, preserving their original granularity to avoid premature abstraction.

**Step 2: Consolidation and normalization.** We normalized terminology and grouped semantically similar criteria while retaining lower-level distinctions. For instance, criteria associated

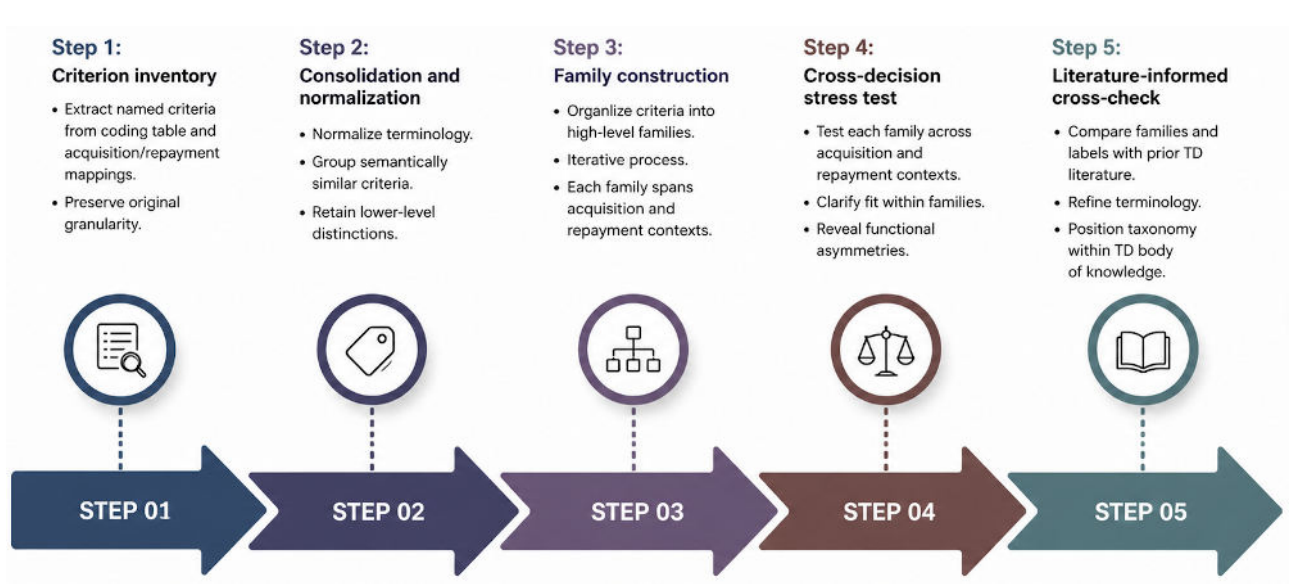

**Figure 1: Overview of the taxonomy process.**

with users, clients, sponsors, and market value were grouped under stakeholder-facing value.

**Step 3: Family construction.** We organized criteria into high-level families through an iterative process, ensuring each family covered both acquisition and repayment contexts and represented a coherent concern.

**Step 4: Cross-decision stress test.** We checked whether each family could classify criteria from both acquisition and repayment contexts. This clarified how criteria fit within each family and exposed functional asymmetries between decision moments.

**Step 5: Literature-informed cross-check.** We compared the emerging families and labels with prior TD literature [10, 15, 17, 22]. This step supported terminology refinement and positioned the taxonomy relative to established TD concepts, types, management activities, and prioritization schemas.

## 3.5 Rigor and Traceability

We adopted four strategies to support rigor. First, we preserved traceability from families to lower-level criteria and, when possible, to original code identifiers. Second, we defined families by their decision meaning, not only by surface similarity. Third, we used acquisition and repayment mappings as a cross-decision check, requiring each family to be meaningful across both decision moments. Fourth, we used prior TD literature as a post-analysis conceptual cross-check to refine terminology while preserving the empirical origin of the categories.

Because the taxonomy involves interpretive abstraction, the analytical artifacts were used to document how we moved from participant accounts to taxonomic families. The coding table links criteria to interview-derived codes, while mapping documents link criteria to patterns and themes.

## 3.6 Data Availability

The anonymized data, including the interview guide and transcripts, coding table, acquisition and repayment criteria mappings, and pattern descriptions, have been deposited in the Zenodo repository and are available at https://shortlink.uk/1o-DN. A README file in the repository documents the package structure, the transcription and anonymization steps, and how each artifact (transcripts, codes, patterns, and interview guide) maps to the analysis reported in this paper.

# 4 Findings

This section presents our findings by research question. Section 4.1 identifies stakeholder criteria across TD decision-making contexts. Section 4.2 presents the proposed taxonomy and conceptual model. Section 4.3 discusses the structural asymmetries between acquisition and repayment criteria. We include short illustrative quotations from participants, translated from Portuguese, to preserve the connection between the taxonomy and the empirical material. The quotations illustrate representative patterns rather than exhaustively documenting each criterion.

## 4.1 RQ1: Stakeholder Criteria Across TD Decision-Making Contexts

As a result of the criterion inventory and consolidation steps described in Section 3.4, our analysis identified several criteria practitioners use to reason about TD decisions: (i) visible business concerns, such as client satisfaction, stakeholder expectations, time-to-market, cost, and alignment with business goals; (ii) technical concerns, such as architecture, system stability, test debt, code readability, bug severity, product quality, and maintainability; and (iii) socio-organizational concerns, such as team alignment, developer satisfaction, organizational constraints, process maturity, negotiation, and stakeholder influence.

A first observation is that TD criteria are not limited to properties of the debt item itself. Practitioners also use criteria related to who is affected, who has authority, whether the issue is visible, whether the team has time to act, and whether the organization can legitimize the work. This shows that TD decision-making criteria are socio-technical, combining technical outcomes with their interpretation within social, organizational, and economic contexts.

A second observation is that criteria appear at different levels of abstraction. Some criteria refer to concrete technical properties, such as bugs or system stability. Others relate to decision conditions, such as urgency, deadline pressure, operational slack, or resource availability. Others point to legitimation mechanisms, such as evidence-based justification, internal consensus, negotiation, or business prioritization. These differences matter: a criterion may identify what is at stake, justify urgency, explain delay, or make the decision acceptable to stakeholders.

A third observation is that criteria are often shared across stakeholder groups, but interpreted differently. For instance, risk may mean system failure for a technical lead, client dissatisfaction for a product manager, or delay for a business stakeholder. Similarly, quality may mean user-visible product quality, internal code quality, process quality, or future maintainability. This suggests the same label can hide different interpretations of value.

Taken together, these observations consolidate prior socio-technical accounts of TD decision-making as stakeholder criteria.[1]

> **Finding 1.** TD decision-making criteria form a heterogeneous socio-technical set that includes technical, business, temporal, organizational, epistemic, and human concerns.

[1] Finding 1 consolidates prior socio-technical observations in TD research rather than claiming that criterion heterogeneity itself is new; it establishes the empirical basis for the taxonomy developed in RQ2.

**Finding 2.** Criteria differ not only by topic but also by decision function: some criteria define value, some trigger action, some justify delay, and others legitimate intervention.

## 4.2 RQ2: A Practitioner-Informed Taxonomy Represented as a Conceptual Model

RQ2 asked how stakeholder criteria can be organized into a taxonomy for TD decision-making. We answer by presenting a practitioner-informed taxonomy represented as a conceptual model. The taxonomy organizes stakeholder criteria into six taxonomic families, while the conceptual model shows how these families operate across TD acquisition and repayment.

The conceptual model was derived from our five-step taxonomy process described in Section 3.4. We extracted criteria from the coding table and acquisition/repayment mappings, consolidated semantically related criteria, organized them into families, tested these families across both decision contexts, and cross-checked the resulting structure against prior TD literature to refine terminology and conceptual boundaries. This cross-check showed that the model relates to established TD concerns such as TD types, management activities, prioritization schemas, and payment practices [9, 10, 15, 17, 22], but organizes them from a different analytical angle: the criteria stakeholders mobilize when authorizing, postponing, negotiating, or legitimizing TD-related decisions.

Figure 2 presents the conceptual model derived from our taxonomy construction process. The model represents TD decision-making as a stakeholder-mediated process rather than as a direct reaction to a technical debt item. It is organized into four conceptual components: the TD decision context, the six taxonomic families of stakeholder criteria, the mediation mechanisms through which criteria are mobilized, and the resulting decision function.

As shown in Figure 2(a), the TD decision context creates the need for stakeholders to decide about either TD acquisition or TD repayment. In acquisition, the decision concerns whether an imperfect solution, shortcut, or compromise can be accepted under current constraints. In repayment, the decision concerns whether an intervention in existing debt should be prioritized, delayed, negotiated, or rejected.

Figure 2(b) illustrates the six taxonomic families that organize the stakeholder criteria mobilized during the decision. The taxonomy consists of: STAKEHOLDER-FACING VALUE, DELIVERY AND RESOURCE PRESSURE, TECHNICAL INTEGRITY AND SYSTEMIC RISK, DECISION BASIS AND EPISTEMIC STYLE, GOVERNANCE AND LEGITIMATION, and HUMAN AND TEAM SUSTAINABILITY. Each family captures a different dimension of the decision situation. These families do not represent sequential stages. Instead, they organize the concerns stakeholders may invoke when evaluating a TD decision.

Figure 2(c) shows the mediation mechanisms through which stakeholders use criteria from these families: interpretation, translation, and legitimation. Interpretation refers to how stakeholders assign meaning to a criterion in a specific decision situation. For example, DELIVERY AND RESOURCE PRESSURE may be interpreted as an unavoidable deadline in acquisition, but as lack of available capacity in repayment. Translation refers to how stakeholders reformulate a concern so that it can be understood across roles. For instance, limited team capacity may be translated into delivery risk for managers, while recurring defects may be translated into business continuity concerns for clients or sponsors. Legitimation refers to how a criterion becomes acceptable as a basis for action, such as when deadline pressure justifies a shortcut or when evidence of recurring incidents justifies repayment. In this sense, the families provide the vocabulary through which stakeholders reason about the decision, while interpretation, translation, and legitimation describe how this vocabulary is used.

Finally, Figure 2(d) represents the resulting decision function produced by this mediation. The model does not imply that criteria mechanically determine a positive decision. Instead, criteria support a decision function that may either enable or block action. In acquisition, this function is permissive: stakeholders may accept or reject an imperfect solution depending on whether the mobilized criteria make the compromise acceptable. In repayment, this function is authorizing: stakeholders may approve, postpone, or reject intervention depending on whether the mobilized criteria make repayment visible, legitimate, and worth allocating resources to.

Table 2 summarizes the taxonomy and complements the model by detailing each taxonomic family, the type of criteria it captures, example criteria, and its typical function in acquisition and repayment contexts. We organized the criteria into six taxonomic families: (i) STAKEHOLDER-FACING VALUE; (ii) DELIVERY AND RESOURCE PRESSURE; (iii) TECHNICAL INTEGRITY AND SYSTEMIC RISK; (iv) DECISION BASIS AND EPISTEMIC STYLE; (v) GOVERNANCE AND LEGITIMATION; and (vi) HUMAN AND TEAM SUSTAINABILITY.

The connection between the families and the mediation mechanisms can be seen in how a family is used in context. For example, DELIVERY AND RESOURCE PRESSURE can be interpreted as deadline pressure, translated into the need to protect delivery commitments, and legitimized as a reason to accept a shortcut in acquisition. In repayment, the same family can be interpreted as lack of capacity, translated into competition with feature work, and used to legitimize postponing intervention. Similarly, STAKEHOLDER-FACING VALUE can legitimize acquisition when responsiveness to a client is needed and can legitimize repayment when TD becomes visible as user dissatisfaction. Thus, Table 2 operationalizes the conceptual model by showing how each family connects to different decision functions across acquisition and repayment.

#### 4.2.1 *Stakeholder-facing value.*

This family includes criteria that evaluate TD decisions based on their effects on clients, users, sponsors, and other external stakeholders. Examples include client satisfaction, user impact, stakeholder influence, market pressure, time-to-market, user value perception, and perceived business value.

This family captures cases where TD decisions are justified by preserving or enhancing visible value. In acquisition, it can authorize shortcuts when they enable responsiveness, visible progress, client satisfaction, or timely delivery. In repayment, it can authorize intervention when TD consequences become visible to users, clients, or sponsors.

> *"Sometimes the client needs an urgent feature, and you have to find a middle ground."* — P11, Tech Lead

The repayment side appeared when internal technical concerns became visible to users or clients: *"If it is affecting the client, then we fix it right away"* (P3, Developer).

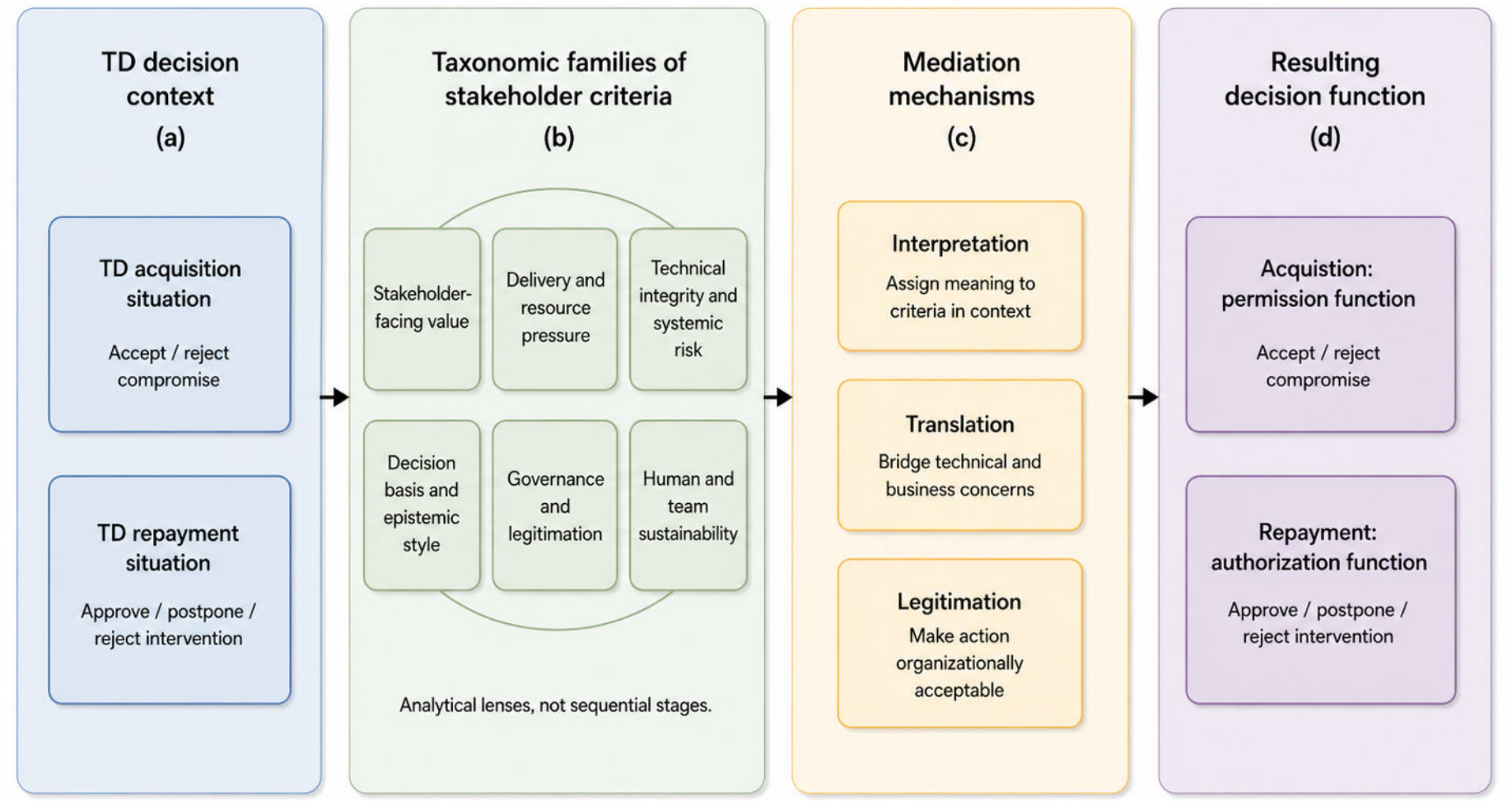

**Figure 2: Conceptual model of stakeholder criteria mobilization in TD decision-making. The model represents TD decision-making as a stakeholder-mediated process composed of four elements: (a) the TD decision context, (b) the taxonomic families of stakeholder criteria, (c) the mediation mechanisms through which criteria are mobilized, and (d) the resulting decision function.**

**Table 2: Taxonomy of stakeholder criteria for TD decision-making.**

| Taxonomic family | What it captures | Example criteria | Typical acquisition function | Typical repayment function |
|---|---|---|---|---|
| Stakeholder-facing value | Effects on clients, users, sponsors, market expectations, and perceived external value | Client satisfaction; user impact; stakeholder influence; stakeholder satisfaction; time-to-market; user value perception | Makes shortcuts acceptable when they deliver visible value | Makes repayment acceptable when TD harms visible value |
| Delivery and resource pressure | Time, cost, capacity, workload, planning constraints, and operational windows | Deadline pressure; time; cost; resource availability; productivity; operational slack; release planning | Accelerates shortcuts and narrows feasible choices | Delays, schedules, or enables intervention |
| Technical integrity and systemic risk | Internal system health, maintainability, architecture, stability, and risk exposure | Architecture; system stability; product quality; process quality; test debt; bugs; security; criticality; maintainability | Defines what is being compromised | Defines what must be protected or restored |
| Decision basis and epistemic style | How stakeholders know, estimate, or justify that a decision is appropriate | Evidence; intuition; prior experience; severity; ROI; cost-benefit reasoning; recurring incidents | Supports quick judgment under uncertainty | Makes invisible debt legible and defensible |
| Governance and legitimation | How TD decisions become organizationally acceptable or authorized | Business prioritization; consensus; negotiation; persuasion; organizational constraints; process maturity; backlog integration | Frames debt as an acceptable trade-off | Frames repayment as legitimate work |
| Human and team sustainability | Effects on morale, cognitive load, autonomy, ownership, learning, and work friction | Developer satisfaction; team alignment; ownership; fear of breaking things; support burden; team morale | Preserves flow and immediate momentum | Reduces cognitive burden and team friction |

> **Finding 3.** Stakeholder-facing value is a cross-cutting family, but its decision function changes across moments: it often legitimizes acquisition through responsiveness and legitimizes repayment through visible dissatisfaction or user impact.

*4.2.2 Delivery and resource pressure.* This family includes criteria about time, deadlines, speed, cost, capacity, workload, resource availability, and delivery planning. Examples include deadline pressure, time, cost, resource availability, productivity, release planning, time-to-market, operational lulls, and sprint capacity.

This family describes the project conditions under which TD decisions become plausible. In acquisition, these often make shortcuts appear necessary. In repayment, the same factors constrain action; debt may be recognized but postponed due to capacity limits, feature delivery taking priority, or the absence of an operational lull.

Participants described acquisition as a response to compressed time: *"We have to deliver quickly, so we end up doing what we can"* (P1, Developer). Conversely, repayment was often associated with temporal slack: *"We pay it when we have more time, when there is a break"* (P3, Developer).

**Finding 4.** DELIVERY AND RESOURCE PRESSURE acts as a temporal gate: it accelerates TD acquisition but frequently postpones TD repayment.

#### 4.2.3 *Technical integrity and systemic risk.*

This family includes criteria that evaluate the internal health of the software system, such as architecture, stability, product and process quality, code complexity, test debt, bug severity, infrastructure reliability, security, maintainability, scalability, criticality, and risk.

This family is closest to traditional TD reasoning. However, technical integrity does not automatically dominate decisions; it often competes with delivery pressure, client expectations, cost constraints, and organizational priorities. It becomes decisive when stakeholders translate it into risk, business continuity, user impact, or future development capacity.

The technical rationale was most explicit when participants framed TD as a threat to the system core: *"There are things that need to be done to avoid critical risks that could lead us to redo everything"* (P11, Tech Lead). Developers also associated repayment with maintainability: *"When the code is too messy, we need to fix it so we can maintain it later"* (P1, Developer).

**Finding 5.** TECHNICAL INTEGRITY AND SYSTEMIC RISK is central to TD decision-making, but rarely acts alone; it becomes influential when translated into risk, user impact, cost, or business continuity.

#### 4.2.4 *Decision basis and epistemic style.*

This family captures how stakeholders know or claim to know that a TD decision is appropriate. Examples include evidence-based justification, professional intuition, prior experience, previous TD impact, informal cost-benefit estimation, perceived ROI, recurring incidents, severity assessment, team maturity, and requirement criticality.

This family shows that TD criteria are not always applied through formal measurement. Practitioners frequently rely on experience, intuition, local knowledge, or visible signals. At the same time, evidence and artifacts become important when TD must be made legible to other stakeholders. Repayment may require evidence because internal quality problems are less visible than features or user-facing defects; acquisition may rely more on urgency, experience, and pragmatic judgment under time constraints.

This contrast appeared clearly in the data. One developer described TD timing as experiential judgment: *"We feel when it is time to pay or when we can leave it for later"* (P1, Developer). By contrast, when repayment needed approval, evidence became persuasive: *"If you show with data that the problem is recurring, it is easier to get approval to fix it"* (P8, Requirements Analyst).

**Finding 6.** DECISION BASIS AND EPISTEMIC STYLE shows that TD criteria are frequently applied through mixed epistemic styles: formal evidence, process artifacts, professional intuition, and experiential heuristics coexist in the same decision space.

#### 4.2.5 *Governance, negotiation, and legitimation.*

This family includes criteria and mechanisms that determine whether a TD decision becomes organizationally acceptable. Examples include business prioritization, alignment with business goals, organizational constraints, process maturity, internal consensus, negotiation, persuasion, stakeholder approval, backlog integration, and communication across roles.

This family shows that TD decisions often require legitimation before they can be enacted. Acquisition may require legitimizing a shortcut as necessary or acceptable. Repayment may require legitimizing maintenance work as valuable enough to compete with feature delivery. In both cases, TD decisions depend not only on what is technically appropriate but also on what the organization recognizes as a valid use of time and resources.

Participants described this legitimation work as collective and negotiated. A developer noted: *"We discuss these technical debt issues in planning, with the whole team, including the PO"* (P1, Developer). In other cases, legitimation involved external negotiation: *"We managed to convince the client of its importance, and the client took on that cost, which is really not cheap"* (P11, Tech Lead).

**Finding 7.** GOVERNANCE AND LEGITIMATION operates as a boundary mechanism that transforms TD from a technical concern into an authorized organizational decision.

#### 4.2.6 *Human and team sustainability.*

This family includes criteria related to morale, autonomy, team alignment, developer satisfaction, learning, ownership, fear of breaking things, support burden, cognitive relief, and development culture.

This family shows that TD decisions are shaped by the human cost of software work. Stakeholders may acquire debt to preserve delivery flow or immediate momentum. They may repay debt to reduce cognitive burden, improve morale, prevent burnout, restore developer confidence, or improve the experience of working with the codebase.

Participants connected TD decisions to everyday development sustainability. A tech lead emphasized developer satisfaction: *"The developer has to be satisfied with the process. This is crucial because it makes their work much easier"* (P11, Tech Lead). Similarly, a requirements analyst described repayment as reducing repeated friction: *"In the long run, the team gains efficiency because it no longer has to work around the same problem several times"* (P8, Requirements Analyst).

**Finding 8.** HUMAN AND TEAM SUSTAINABILITY reveals TD decision-making as a labor and coordination problem, not only a technical or economic one.

## 4.3 RQ3: Structural Asymmetries Between Acquisition and Repayment Criteria

The taxonomy reveals that TD acquisition and TD repayment are not symmetrical decisions. They mobilize overlapping criterion families, but these families perform different decision functions. Acquisition criteria tend to operate as permission mechanisms: they justify why a shortcut is acceptable. Repayment criteria tend to operate as authorization mechanisms: they justify allocating time and resources to address existing debt.

#### 4.3.1 *Acquisition Criteria Authorize Shortcuts.*

In acquisition, stakeholders use criteria to justify action under constraint. A shortcut may be acceptable if the client needs a feature, the deadline is close, the market window is narrow, the current risk seems manageable, or the team believes it can address the issue later. In these situations, criteria preserve momentum and frame the compromise as rational, necessary, or temporary.

This does not mean acquisition decisions are careless. Rather, acquisition criteria often convert constraint into legitimacy. Delivery and resource pressure narrows feasible alternatives. Stakeholder-facing value makes the shortcut meaningful. Decision basis and epistemic style provides confidence that the compromise is acceptable. Governance and legitimation makes the decision organizationally defensible.

#### 4.3.2 *Repayment Criteria Authorize Intervention.*

In repayment, stakeholders use criteria to justify interrupting or slowing feature delivery. Repayment competes with visible work, so its criteria must make the cost of inaction visible. A repayment effort becomes legitimate when debt affects users, causes recurring incidents, slows developers, threatens system stability, creates risk, or is recognized by a formal process.

This gives repayment a different rhetorical burden. Acquisition can be justified by immediate value, whereas repayment often requires demonstrating that hidden or delayed costs warrant present attention. Technical integrity and systemic risk alone may not be enough; stakeholders often need to connect it to risk, user impact, cost, or team sustainability.

#### 4.3.3 *Shared Families, Different Functions.*

Table 3 summarizes the asymmetry. The same family may authorize different actions at different times. For example, stakeholder-facing value can justify acquiring TD to quickly satisfy a client, but it can also justify repayment when TD harms the user experience. delivery and resource pressure can justify acquiring TD under a deadline, but can postpone repayment when there is no time available. technical integrity and systemic risk can be compromised during acquisition, but becomes a basis for intervention during repayment.

**Finding 9.** Acquisition and repayment share criterion families, but differ in decision function: acquisition criteria tend to permit shortcuts, while repayment criteria tend to authorize intervention.

**Finding 10.** The taxonomy exposes a temporal asymmetry: acquisition is usually justified by immediate delivery value, whereas repayment requires the delayed consequences of TD to become visible, negotiable, or institutionally recognized.

## 5 Discussion

Our study shows that TD decisions are shaped by multiple criteria whose meanings and functions shift across decision moments. By representing the taxonomy as a conceptual model, we show how stakeholder criteria are mobilized, interpreted, translated, and legitimized before becoming actionable in TD acquisition or repayment decisions.

**Table 3: Structural asymmetries between TD acquisition and repayment.**

| Family | In acquisition | In repayment |
|---|---|---|
| Stakeholder-facing value | Responsiveness makes debt acceptable | Visible dissatisfaction makes repayment legitimate |
| Delivery and resource pressure | Deadlines accelerate shortcuts | Lack of capacity delays repayment |
| Technical integrity | Defines what is knowingly compromised | Defines what must be restored or protected |
| Decision basis | Supports judgment under uncertainty | Makes hidden debt defensible through evidence |
| Governance and legitimation | Frames debt as an acceptable trade-off | Frames maintenance as authorized work |
| Human and team sustainability | Preserves immediate flow | Reduces future friction and cognitive load |

### 5.1 From Criterion Heterogeneity to Criterion Function

Our taxonomy shifts attention from which criteria exist to what criteria do. Many criteria identified in RQ1 could be reported as a flat list: time, cost, client satisfaction, system stability, risk, product quality, team alignment, stakeholder influence, and so on. While such lists are useful and match elements frequently mapped in TD studies and reviews [9, 14, 22], they obscure an important empirical pattern: criteria are not equivalent decision units. Some express value; others express constraint, technical concern, evidence, or authorization.

This distinction matters because the same criterion can participate in different decision mechanisms. Time may refer to deadline pressure, delivery speed, future effort, time-to-market, or maintenance slack. Quality may refer to user-visible adequacy, internal code quality, process quality, or architectural sustainability. Client satisfaction may justify taking on TD or trigger paying it back. Therefore, the taxonomy proposed in RQ2 does not simply group similar labels, as prior structural classifications of TD types [1, 15] do. It organizes criteria according to the role they play in making TD decisions actionable. The conceptual model extends this contribution by showing that criteria become actionable through stakeholder interpretation, translation across perspectives, and organizational legitimation.

This helps explain why TD management is difficult in practice. Teams may agree on a criterion but disagree on its function. A manager and a developer may both mention risk, but one may refer to delivery commitments and the other to system evolution. A product and a technical role may both mention quality, but one may mean user-visible adequacy and the other maintainability. The taxonomy clarifies these differences by separating stakeholder value, technical integrity, epistemic basis, governance, and human sustainability. The model further suggests that disagreement is not merely terminological; it is mediated by how stakeholders assign meaning to criteria and translate them into concerns that other groups recognize as legitimate.

## 5.2 The Asymmetry Between Permission and Authorization

Our findings show that the core distinction between acquisition and repayment is the reasoning: permission versus authorization. In acquisition, criteria act as permission structures: they rationalize a shortcut as responsive, necessary, temporary, or proportionate. Delivery pressure enables speed, client value enables compromise, informal judgment permits action under uncertainty, and governance frames the shortcut as an acceptable trade-off. In repayment, criteria act as authorization structures: they justify why the organization should allocate current resources to a past commitment. Repayment decisions require more than acknowledging a debt item; they demand demonstrating that the debt is visible, risky, costly, painful, strategically relevant, or procedurally legitimate. This interpretation builds on prior work showing that TD repayment is hindered by low priority, lack of organizational interest, short-term goals, costs, and time constraints [9], as well as work indicating that architectural and organizational factors influence how TD accumulates and is addressed over time [18, 19].

This asymmetry explains why TD is often easier to acquire, even when known to be technically imperfect, but harder to repay, even when known to be undesirable. Acquisition benefits from immediacy—delivery, features, clients, deadlines, and momentum. Repayment suffers from invisibility, competes with new work, and converts delayed consequences into present claims. The taxonomy captures this paradox by showing that acquisition and repayment share families, not functions. The conceptual model makes this asymmetry explicit: in acquisition, criteria help authorize compromise; in repayment, they help authorize intervention.

## 5.3 Visibility, Temporality, and Authority as Cross-Cutting Tensions

Overall, we observed three tensions operating within the taxonomic families: visibility, temporality, and authority. Although prior TD research [15, 22] has shown that TD management requires identification, communication, prioritization, and repayment, our findings suggest that these activities depend not only on recognizing TD items but also on making the criteria for action visible, timely, and legitimate.

**Visibility** concerns whether a criterion can be perceived by the stakeholders who control the decision. Stakeholder-facing value is highly visible because clients, users, and sponsors can often notice missing functionality, frustration, or delivery delays. Technical integrity is less visible until it appears as defects, instability, slow development, or user impact. This explains why repayment often requires evidence-based justification: technical debt must be translated from internal friction into visible organizational relevance.

**Temporality** concerns whether a criterion points to immediate or delayed consequences. Acquisition criteria are frequently immediate: a deadline, client request, release, sprint commitment, or market opportunity. Repayment criteria often concern delayed consequences: future maintenance, accumulated instability, developer burden, recurring incidents, or long-term risk. TD decisions therefore involve a temporal imbalance: the present is easier to defend than the future, especially when the future competes with visible delivery pressure.

**Authority** concerns who has the legitimacy to make the criterion count. A developer may recognize architectural degradation, but this does not guarantee repayment. A product manager may recognize user dissatisfaction, but this does not identify the technical intervention needed. Process maturity may enable repayment or create bureaucratic friction. Governance and legitimation criteria are therefore not peripheral; they determine whether other criteria can become actionable.

These tensions show why the taxonomy is more than a classification tool. It allows analysis of how criteria move across stakeholders and decision moments. A criterion is powerful when it is visible enough, urgent, and backed by the right authority. Conversely, even correct criteria may remain weak if they are invisible, future-focused, or unsupported. This is where the conceptual model adds explanatory value: it shows that criteria must pass through mediation mechanisms before they can support an authorized decision function.

## 5.4 Relation to Prior TD Organizing Artifacts

Our results complement prior TD organizing artifacts. Existing taxonomies and mappings clarify TD types, TD management activities, tools, strategies, and payment practices [9, 15, 22]. Other studies propose prioritization schemas and management diagrams that distinguish technical and nontechnical perspectives or organize practices, impediments, decision factors, and actions [10, 17]. Our contribution differs by changing the unit of organization: instead of classifying TD items, management activities, or payment practices, we organize stakeholder criteria and their decision functions across acquisition and repayment.

The literature-informed cross-check helped refine terminology and avoid positioning the taxonomy as an isolated or self-contained classification. However, the resulting model remains empirically grounded: its families were derived from interview-based artifacts and then aligned with prior TD literature. Thus, the model should be read as a practitioner-informed conceptual organization of TD decision criteria, not as a universal theory of all TD decision-making.

## 5.5 Implications for Research and Practice

For research, the taxonomy and conceptual model suggest that future TD studies should avoid treating criteria as flat factor lists. Researchers should specify each criterion's function: whether it defines value, creates pressure, identifies technical risk, provides evidence, legitimizes action, or protects team sustainability. This improves empirical comparability and complements existing TD classifications by shifting the focus from types or repayment practices to the stakeholder criteria used to inform TD decisions [1, 9]. The model also suggests that future studies should examine mediation mechanisms explicitly, especially how stakeholders interpret criteria, translate technical concerns into business-relevant terms, and legitimate TD-related decisions.

For practice, the taxonomy can support more balanced TD discussions during planning, retrospectives, debt reviews, or architectural discussions. When acquiring TD, teams can ask whether a decision is justified only by delivery pressure or whether technical integrity and future repayment conditions have also been considered. When proposing repayment, teams can ask whether the case is framed only as internal quality or also translated into stakeholder-facing value, risk reduction, business continuity, or team sustainability.

The taxonomy suggests TD repayment is not simply a technical backlog problem. If repayment needs authorization, teams must make debt visible, link it to value, and create clear moments for action. Examples include debt review rituals, evidence templates, backlog labels for criterion families, or lightweight decision records explaining why debt was acquired and what would justify repayment. The conceptual model can also support facilitation by helping teams distinguish whether they are accepting a compromise, postponing action, or legitimizing an intervention.

## 6 Threats to Validity

This section discusses the main threats to the validity of our study and the mitigation strategies we adopted, following standard validity dimensions in empirical software engineering research.

**Construct and Internal Validity.** A primary threat concerns the definition and interpretation of the central construct of this paper: the stakeholder criterion. Criteria, factors, motivations, causes, consequences, and contextual conditions often overlap in TD research. To mitigate this threat, we defined stakeholder criteria based on their decision function rather than only on their topic. For example, time was interpreted contextually as deadline pressure, time-to-market, future effort, or available slack.

Because the taxonomy and conceptual model were constructed through qualitative interpretation, researcher bias may have influenced how criteria were consolidated and assigned to families. We mitigated this threat by using multiple analytical artifacts, including coding tables, acquisition and repayment mappings, and pattern descriptions, to cross-check abstractions. We also applied a cross-decision stress test: each proposed family had to be meaningful across both TD acquisition and repayment, avoiding overfitting the taxonomy to a single decision moment. In addition, we used a literature-informed cross-check to refine terminology and conceptual boundaries without imposing categories a priori. Finally, to reduce translation bias, all Portuguese quotations were translated with attention to their analytical role in the surrounding interpretation.

**External Validity.** Our findings are grounded in interviews with 11 software practitioners working in Brazil. Although participants occupied diverse technical and business roles and had substantial industry experience, the sample does not represent all software domains, organizational settings, or national contexts. Therefore, we do not claim statistical generalization.

To mitigate this threat, we position the taxonomy and conceptual model as empirically grounded analytical artifacts, not as a universal theory of TD decision-making. Their scope is limited to capturing socio-technical patterns observed in our study context. These patterns may be transferable to organizational contexts with similar characteristics, but adaptation and refinement may be needed for substantially different settings, such as safety-critical systems, startups, open-source projects, or organizations with alternative governance structures.

**Reliability and Conclusion Validity.** Taxonomy construction involves interpretive judgment. Other researchers might group some criteria differently, especially when criteria span multiple families. For example, customer dissatisfaction could be interpreted as stakeholder-facing value, as evidence for repayment, or as a governance issue, depending on the decision context. To support reliability and transparency, we defined families, provided example criteria, described decision functions, and documented the construction process in five steps: criterion inventory, consolidation and normalization, family construction, cross-decision stress test, and literature-informed cross-check.

Regarding conclusion validity, we kept our claims aligned with the qualitative nature of the evidence. The taxonomy and conceptual model do not predict whether a team will acquire or repay TD, nor do they provide a complete explanatory theory of all TD decisions. The distinction between *permission mechanisms* in acquisition and *authorization mechanisms* in repayment is presented as an analytical pattern grounded in the empirical material, not as a deterministic model.

## 7 Conclusion

This paper proposed a practitioner-informed taxonomy represented as a conceptual model of stakeholder criteria for Technical Debt decision-making. Based on an interview-based qualitative study with 11 software practitioners in Brazil, we extracted, consolidated, and organized the criteria stakeholders use when reasoning about TD acquisition and repayment. Unlike existing TD taxonomies that classify debt items, our model focuses on the criteria stakeholders use to justify, postpone, negotiate, or authorize TD-related decisions. The taxonomy organizes these criteria into six families, showing that TD decisions involve not only technical or economic trade-offs, but also stakeholder-facing value, delivery and resource pressure, technical integrity, epistemic basis, governance, and human/team sustainability.

Our findings show that acquisition and repayment share criterion families but differ in decision function: acquisition criteria tend to permit shortcuts under current constraints, whereas repayment criteria tend to authorize the commitment of resources to address existing debt. The conceptual model further shows that these criteria become actionable through stakeholder interpretation, translation across technical and business perspectives, and organizational legitimation. By providing a cross-decision taxonomy and conceptual model, this paper offers an empirically grounded artifact for comparing TD decisions and supporting more explicit TD discussions in practice. Future work may further evaluate the taxonomy in new settings and explore how specific criteria influence TD-related activities over time.

## References

[1] Nicolli S. R. Alves, Leilane F. Ribeiro, Vivyane Caires, Thiago S. Mendes, and Rodrigo O. Spínola. 2014. Towards an Ontology of Terms on Technical Debt. In *Proceedings of the 2014 Sixth International Workshop on Managing Technical Debt (MTD '14)*. IEEE Computer Society, USA, 1–7. doi:10.1109/MTD.2014.9

[2] Anonymous. 2026. Anonymized for double-blind review. *Anonymized journal* (2026). Details omitted for double-blind review.

[3] Kenneth D. Bailey. 1994. *Typologies and Taxonomies: An Introduction to Classification Techniques.* Number 102 in Quantitative Applications in the Social Sciences. SAGE Publications, Thousand Oaks, CA. doi:10.4135/9781412986397

[4] Christoph Becker, Fabian Fagerholm, Rahul Mohanani, and Alexander Chatzigeorgiou. 2019. Temporal discounting in technical debt: how do software practitioners discount the future?. In *Proceedings of the Second International Conference on Technical Debt* (Montreal, Quebec, Canada) *(TechDebt '19)*. IEEE Press, 23–32. doi:10.1109/TechDebt.2019.00011

[5] João Paulo Biazotto, Daniel Feitosa, Paris Avgeriou, and Elisa Yumi Nakagawa. 2025. Understanding practitioners' reasoning and requirements for efficient tool support in technical debt management. *Empirical Software Engineering* 30, 5 (2025), 134. doi:10.1007/s10664-025-10691-5

[6] Ward Cunningham. 1992. The WyCash portfolio management system. In *Addendum to the Proceedings on Object-Oriented Programming Systems, Languages, and Applications (Addendum)* (Vancouver, British Columbia, Canada) *(OOPSLA '92)*. Association for Computing Machinery, New York, NY, USA, 29–30. doi:10.1145/157709.157715

[7] Carlos Fernández-Sánchez, Juan Garbajosa, and Agustín Yagüe. 2015. A framework to aid in decision making for technical debt management. In *2015 IEEE 7th International Workshop on Managing Technical Debt (MTD)*. 69–76. doi:10.1109/MTD.2015.7332628

[8] Carlos Fernández-Sánchez, Juan Garbajosa, Agustín Yagüe, and Jennifer Perez. 2017. Identification and analysis of the elements required to manage technical debt by means of a systematic mapping study. *Journal of Systems and Software* 124 (2017), 22–38. doi:10.1016/j.jss.2016.10.018

[9] Sávio Freire, Nicolli Rios, Boris Gutierrez, Darío Torres, Manoel Mendonça, Clemente Izurieta, Carolyn Seaman, and Rodrigo O. Spínola. 2020. Surveying Software Practitioners on Technical Debt Payment Practices and Reasons for not Paying off Debt Items. In *Proceedings of the 24th International Conference on Evaluation and Assessment in Software Engineering* (Trondheim, Norway) *(EASE '20)*. Association for Computing Machinery, New York, NY, USA, 210–219. doi:10.1145/3383219.3383241

[10] Sávio Freire, Nicolli Rios, Boris Pérez, Camilo Castellanos, Darío Correal, Robert Ramač, Vladimir Mandić, Nebojša Taušan, Alexia Pacheco, Gustavo López, Manoel Mendonça, Clemente Izurieta, Davide Falessi, Carolyn Seaman, and Rodrigo Spínola. 2021. Pitfalls and Solutions for Technical Debt Management in Agile Software Projects. *IEEE Softw.* 38, 6 (Nov. 2021), 42–49. doi:10.1109/MS.2021.3101990

[11] Tchalisson Brenne S. Gomes, Diogo Alves de Moura Loiola, and Alcemir Rodrigues Santos. 2024. Technical Debt Tools: a Survey and an Empirical Evaluation. 12 (Aug. 2024), 8:1 – 8:16. doi:10.5753/jserd.2024.3591

[12] Yuepu Guo, Rodrigo Oliveira Spínola, and Carolyn Seaman. 2016. Exploring the costs of technical debt management — a case study. *Empirical Softw. Engg.* 21, 1 (Feb. 2016), 159–182. doi:10.1007/s10664-014-9351-7

[13] Philippe Kruchten, Robert L. Nord, and Ipek Ozkaya. 2012. Technical Debt: From Metaphor to Theory and Practice. *IEEE Softw.* 29, 6 (Nov. 2012), 18–21. doi:10.1109/MS.2012.167

[14] Valentina Lenarduzzi, Terese Besker, Davide Taibi, Antonio Martini, and Francesca Arcelli Fontana. 2021. A systematic literature review on Technical Debt prioritization: Strategies, processes, factors, and tools. *Journal of Systems and Software* 171 (2021), 110827. doi:10.1016/j.jss.2020.110827

[15] Zengyang Li, Paris Avgeriou, and Peng Liang. 2015. A systematic mapping study on technical debt and its management. *J. Syst. Softw.* 101, C (March 2015), 193–220. doi:10.1016/j.jss.2014.12.027

[16] Erin Lim, Nitin Taksande, and Carolyn Seaman. 2012. A Balancing Act: What Software Practitioners Have to Say about Technical Debt. *IEEE Softw.* 29, 6 (Nov. 2012), 22–27. doi:10.1109/MS.2012.130

[17] Vladimir Mandić, Nebojša Taušan, Robert Ramač, Sávio Freire, Nicolli Rios, Boris Pérez, Camilo Castellanos, Darío Correal, Alexia Pacheco, Gustavo Lopez, Clemente Izurieta, Davide Falessi, Carolyn Seaman, and Rodrigo Spínola. 2021. Technical and Nontechnical Prioritization Schema for Technical Debt: Voice of TD-Experienced Practitioners. *IEEE Softw.* 38, 6 (Nov. 2021), 50–58. doi:10.1109/MS.2021.3103121

[18] Antonio Martini, Jan Bosch, and Michel Chaudron. 2014. Architecture Technical Debt: Understanding Causes and a Qualitative Model. In *Proceedings of the 2014 40th EUROMICRO Conference on Software Engineering and Advanced Applications (SEAA '14)*. IEEE Computer Society, USA, 85–92. doi:10.1109/SEAA.2014.65

[19] Antonio Martini, Jan Bosch, and Michel Chaudron. 2015. Investigating Architectural Technical Debt accumulation and refactoring over time. *Inf. Softw. Technol.* 67, C (Nov. 2015), 237–253. doi:10.1016/j.infsof.2015.07.005

[20] Robert C. Nickerson, Upkar Varshney, and Jan Muntermann. 2013. A Method for Taxonomy Development and Its Application in Information Systems. *European Journal of Information Systems* 22, 3 (2013), 336–359. doi:10.1057/ejis.2012.26

[21] Leilane Ferreira Ribeiro, Mário André de F. Farias, Manoel Mendonça, and Rodrigo Oliveira Spínola. 2016. Decision Criteria for the Payment of Technical Debt in Software Projects: A Systematic Mapping Study. In *Proceedings of the 18th International Conference on Enterprise Information Systems* (Rome, Italy) *(ICEIS 2016)*. SCITEPRESS - Science and Technology Publications, Lda, Setubal, PRT, 572–579. doi:10.5220/0005914605720579

[22] Nicolli Rios, Manoel Gomes de Mendonça Neto, and Rodrigo Oliveira Spínola. 2018. A tertiary study on technical debt: Types, management strategies, research trends, and base information for practitioners. *Information and Software Technology* 102 (2018), 117–145. doi:10.1016/j.infsof.2018.05.010

[23] Nicolli Rios, Rodrigo Oliveira Spínola, Manoel Mendonça, and Carolyn Seaman. 2018. The most common causes and effects of technical debt: first results from a global family of industrial surveys. In *Proceedings of the 12th ACM/IEEE International Symposium on Empirical Software Engineering and Measurement* (Oulu, Finland) *(ESEM '18)*. Association for Computing Machinery, New York, NY, USA, Article 39, 10 pages. doi:10.1145/3239235.3268917

[24] Marek G. Stochel, Tomasz Borek, Mariusz R. Wawrowski, and Piotr Chołda. 2023. Business-driven technical debt management using Continuous Debt Valuation Approach (CoDVA). *Inf. Softw. Technol.* 164, C (Dec. 2023), 21 pages. doi:10.1016/j.infsof.2023.107333

[25] Edith Tom, AybüKe Aurum, and Richard Vidgen. 2013. An exploration of technical debt. *J. Syst. Softw.* 86, 6 (June 2013), 1498–1516. doi:10.1016/j.jss.2012.12.052

[26] Jesse Yli-Huumo, Andrey Maglyas, and Kari Smolander. 2016. How do software development teams manage technical debt? - An empirical study. *J. Syst. Softw.* 120, C (Oct. 2016), 195–218. doi:10.1016/j.jss.2016.05.018